# Information limit of 15 pm achieved with bright-field ptychography

Haozhi Sha[1,2,3], Jizhe Cui[1,2,3], Wenfeng Yang[1,2,3], Rong Yu[1,2,3,*]

[1]School of Materials Science and Engineering, Tsinghua University, Beijing 100084, China.
[2]MOE Key Laboratory of Advanced Materials, Tsinghua University, Beijing 100084, China.
[3]State Key Laboratory of New Ceramics and Fine Processing, Tsinghua University, Beijing 100084, China.

**Abstract**

It is generally assumed that a high spatial resolution of a microscope requires a large numerical aperture of the imaging lens or detector. In this study, the information limit of 15 pm is achieved in transmission electron microscopy using only the bright-field disk (small numerical aperture) via multislice ptychography. The results indicate that high-frequency information has been encoded in the electrons scattered to low angles due to the multiple scattering of electrons in the objects, making it possible to break the diffraction limit of imaging via bright-field ptychography.

According to Abbe's theory of imaging, the space resolution of a microscope is limited by the diffraction effect of imaging waves [1-5]. The resolution limit can be generally written as $k\lambda/NA$, where $\lambda$ is the wavelength, NA stands for numerical aperture, which is $n\sin\alpha$ with $n$ being the refraction index and $\alpha$ the semi-angle of the objective lens aperture. The factor $k$ varies with resolution criteria [2-5], being 0.61 for Rayleigh resolution, 0.5 for Abbe resolution, and 0.47 for Sparrow resolution, for example. In conventional electron microcopy, Scherzer adopts 0.8 for coherent imaging and 0.6 for incoherent imaging [5].

In scanning transmission electron microscopy (STEM), $\alpha$ is the convergence semi-angle of the probe. Loosely speaking, $k$ is about 1 for coherent modes like bright-field (BF) imaging, and about 0.5 for incoherent modes like high-angle annular dark-field (HAADF) imaging [6-8].

Electron ptychography is a diffractive imaging method that utilizes scattered electrons inside the collection angle $\beta$ of the pixelated detector to retrieve the object information. Consequently, the largest scattering angle or effective aperture is now $\alpha+\beta$, and the diffraction limit of ptychography can be described as $k\lambda/\sin(\alpha+\beta)$. The electrons scattered to the largest angle carry information of the highest spatial frequency that the detector collects. In this way, high-angle scattering has been utilized to improve the resolution of ptychography [9-16]. The information limit reaches 39 pm with a large collection angle $\beta = 3\alpha$, which corresponds to $0.8\lambda/\sin(\alpha+\beta)$ [10].

It has been demonstrated that bringing high-angle scattering wave outside the aperture back to the low frequency region can help to break the diffraction limit of imaging. For example, McCutchen noted that, by putting a stop right against the illuminated object, the low-frequency region of an image will contain contributions from high frequency information in the object due to the convolution process [17]. In optical microscopy, high-angle scattering wave is redirected back by using the multiple scattering of light by disordered media, which can be further utilized to overcome the diffraction limit [18-21]. It is also worth noting that resolution can be enhanced by allowing the iterative ptychography algorithm to recover the diffraction intensity in the frequency region beyond the collection angle of detectors [15,22,23].

Due to the strong interactions between incident electrons and objects, multiple scattering is common in electron microscopy. It usually leads to undesirable effects in diffraction, imaging, and spectroscopy. It is often necessary to develop techniques to mitigate multiple scattering effects. For example, dynamical electron diffraction prevents direct Fourier analysis in crystallography and precession electron diffraction is developed to reduce dynamic diffraction [24,25].

In this work, we show that multiple scattering of electrons can be used in multislice electron ptychography [12,26,27] to extend the information limit in the phase images without the aid of any additional medium or high-angle scattering. The information limit reaches 15 pm, corresponding to $0.44\lambda/\sin(\alpha+\beta)$.

First, we use simulation to demonstrate the effect of multiple scattering in ptychography. Four-dimensional datasets are simulated with the multislice method [28,29] for two thickness values, 0.8 nm and 20 nm, respectively. A perovskite oxide $DyScO_3$ in the [001] zone axis is used as the object. Electron beam of 300 kV and 25-

mrad convergence semi-angle is used to generate simulation datasets, which are divided into low-angle and high-angle groups, with the maximum collection semi-angle of 27 mrad ($\beta \approx \alpha$) and 54 mrad ($\beta \approx 2\alpha$), respectively. The diffraction patterns in all the datasets have 120×120 pixels. Diffraction patterns in both the low-angle and high-angle groups are padded to 120 mrad with zero, resulting in a real space pixel size of 8.2 pm. In order to study only the effects of multiple scattering, the results in FIG. 1 exclude the effects of thermal diffuse scattering and noise. Instead of allowing the iterative algorithm to fill in the extended region in diffraction pattern [15,22,23], we use the conventional Fourier constraint which forces the zero-padded region to zero during optimization.

FIG. 1(a) shows the ptychographic phase images recovered from the simulation datasets of 0.8-nm-thick sample. FIG. 1(b) shows the corresponding Fourier transforms, which are azimuthally averaged. The information limits ($d_{info}$) are labeled. Phase profiles across the atomic columns of O-Sc-O and isolated O are displayed in FIG. 1(c). It is clear that large collection angle increases the information limit, consistent with the observation in Ref. [10]. For very thin samples, most incident electrons are scattered no more than once before reaching the detector. The kinematic approximation is applicable and the maximum scattering angle $\theta_{max}$ of electrons collected by the detector increases with the collection semi-angle $\beta$. As electrons scattered to high angles carry high frequency information, the information limit in the recovered phase images is limited by the collection semi-angle, as shown in FIG. 1 (a), (b) and (c).

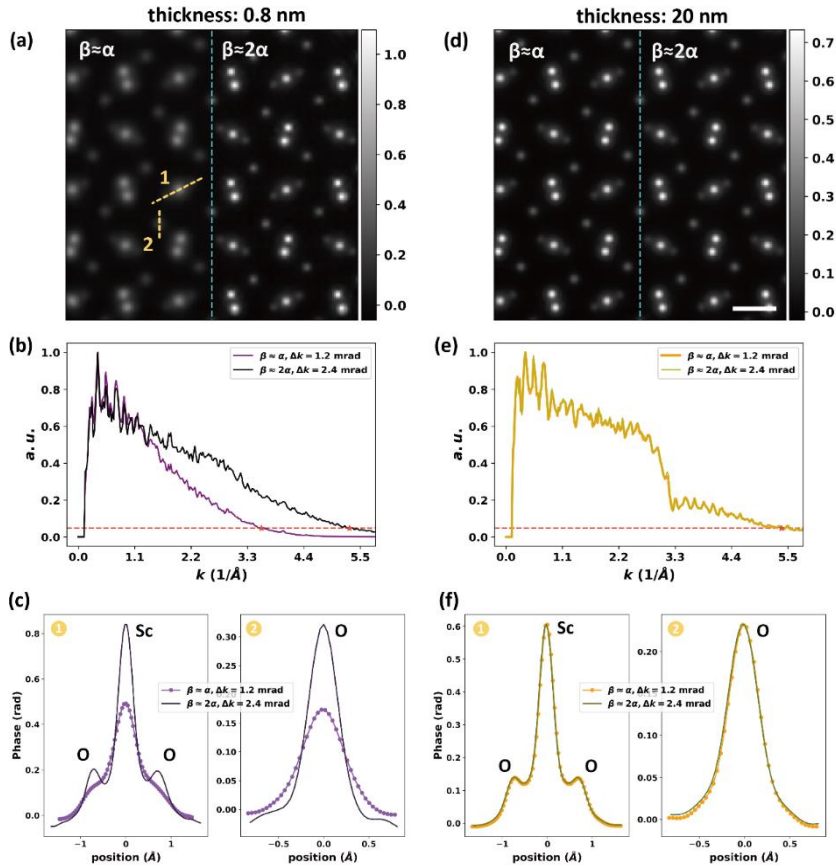

FIG. 1. Resolution improvement by using multiple scattering. Sample thickness of (a, b and c) is

0.8 nm, (d, e and f) 20 nm. (a, d) Comparison of recovered mean phases based on datasets with different collection semi-angle. Each phase image is averaged over all the slices. Collection angles are labeled on the top of each image. Slice thickness is chosen as 0.5 nm during multislice ptychographic reconstruction. Scale bar, 4Å. (b, e) Azimuthal average of the power spectrums corresponding to the phase images shown in (a and d). Information limits are marked with red starts. (c, f) Phase profiles across the lines marked in (a) and (d).

FIG. 1(d) shows the ptychographic phase images recovered from the simulation datasets of 20-nm-thick sample. It is noticed that the phase image for the small collection angle is nearly the same as that for the large collection angle. As shown in FIG. 1(e), the information limit is the same for the small and large collection angles. The broadening widths of atomic columns become nearly identical for the two collection angles, as shown in FIG. 1(f), indicating that the bright-field disc contains sufficient high-frequency information that can be retrieved by multislice ptychography. When the thickness of samples increases, multiple scattering becomes prevalent. The largest scattering angle of an electron is no longer limited by the collection angle. Electrons scattered to high angles can be scattered back and enter a low-angle aperture. Therefore, electrons hit the low-angle area in a detector may also carry high-frequency information. Results recovered from datasets considering the thermal diffuse scattering are shown in FIG. S1, which lead to the same conclusion.

Furthermore, we demonstrate in experiment that the information limit of ptychography can surpass the diffraction limit of $0.5\lambda/\sin(\alpha+\beta)$ by taking advantage of multiple scattering. The specimen of $DyScO_3$ single crystal is 25 nm in thickness. 4D datasets are acquired using an electron microscope pixel array detector (EMPAD) [30] under a high voltage of 300 kV and convergence semi-angle $\alpha$ of 25 mrad. Each diffraction pattern contains 128×128 pixels. Different camera lengths are used to get reciprocal sampling intervals of 0.055, 0.042, 0.026 and 0.021 Å$^{-1}$ (corresponding collection semi-angles $\beta$: 67, 51, 32 and 26 mrad). The results of $\beta$ = 26 mrad (~1.0$\alpha$) and 51 mrad (~2.0$\alpha$) are shown in FIG. 2, those of $\beta$ = 32 mrad (~1.3$\alpha$) and 67 mrad (~2.7$\alpha$) are shown in FIG. S2. The position-averaged CBEDs (PACBEDs) are shown as insets in FIG. 2. For a fair comparison, diffraction patterns in the first three datasets are all zero-padded to 154 mrad to make sure the real space pixel size of the recovered phase image reaches 64 pm. Because of the limit of GPU memory, diffraction patterns with $\beta$ = 26 mrad are only padded to 132 mrad, corresponding to a real space pixel size of 75 pm. The sample is divided into 3-Å slices during reconstruction. The recovered phase images are shown in FIG. 2 (a, b). From the power spectra shown in FIG. 2 (c, d), we can see that the resolution improves with smaller collection angles. For the condition of $\beta \approx 2\alpha$ (FIG. 2 (a) and (c)), the information limit reaches 18 pm, overcoming the limit of $\lambda/(\alpha+\beta)$. Further refining the sampling in reciprocal space promotes the information limit to 15 pm (FIG. 2 (b) and (d)), which corresponds to $0.44\lambda/(\alpha+\beta)$, breaking the diffraction limit in the strictest sense.

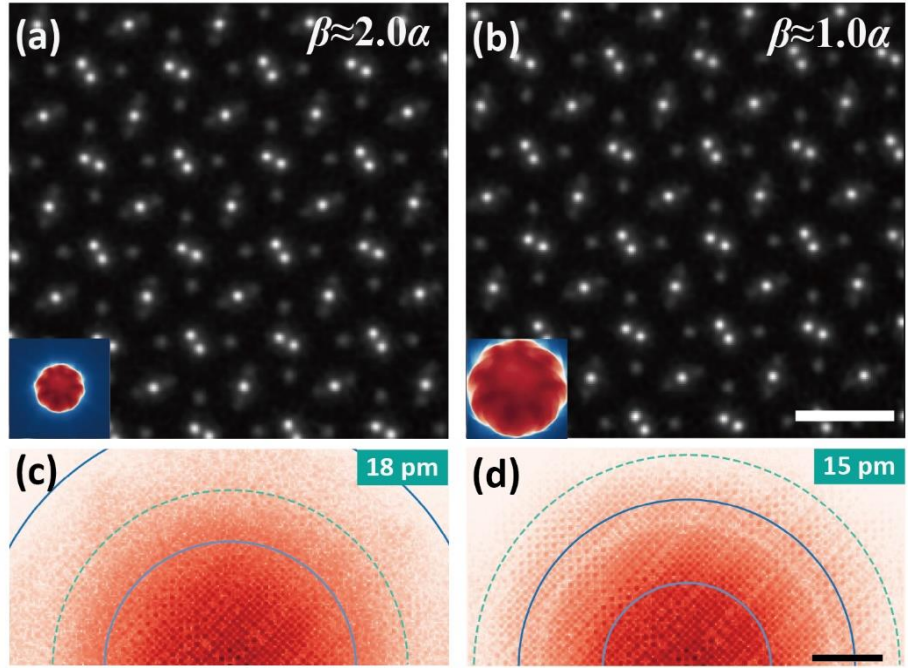

FIG. 2. Experimental results using different collection semi-angles. (a and b) Reconstructed DyScO$_3$ phase images averaged over slices free of the influence of surface damage. Slice thickness is 3 Å. Scale bar, 6 Å. Insets show PACBEDs. Sampling interval in reciprocal space is 0.042 and 0.021 Å$^{-1}$, respectively. (c and d) Corresponding power spectrums of phases shown in (a and b). Information limits are marked with cyan dashed lines and labeled on the top left. Diffraction limits corresponding to $\lambda/(\alpha+\beta)$ and $0.5\lambda/(\alpha+\beta)$ are marked with light and dark blue solid lines, respectively. Scale bar, 2 Å$^{-1}$.

The experimental results confirm the enhancement effect of multiple scattering to the ptychographic resolution. For a specific pixelated detector, decreasing $\Delta k$ results in smaller collection angle. However, as the simulation predicted, the negative effect of smaller $\beta$ is compensated by multiple scattering. With sufficient electron dose, fine reciprocal sampling further makes the high-frequency information generated by multiple scattering in diffraction patterns be better decoded, prompting the resolution of recovered phase images. Consequently, even under the condition of $\beta \approx \alpha$, the information limit reaches 15 pm.

To analyze the relationship between the magnitude of multiple scattering and the resolution of ptychographic phases in detail, we performed extensive multislice simulations of CBEDs and ptychographic reconstructions for samples with a series of thicknesses. The full widths at 80% of the maximum (FW80Ms) of oxygen columns were measured and used to assess the resolution of ptychographic phases (FIG. 3). FW80Ms of oxygen columns in the ptychographic phase images is shown with triangles and diamonds in FIG. 3 for small and large collection angles, respectively. Reconstruction result in FIG. 3(a) is based on noise-free datasets and result in FIG. 3(b) is based on datasets under the same electron dose value with the experiment ($4\times10^6$ e/Å$^2$). As the number of pixels of the detector is fixed in experiments, the dataset with large collection angle means low sampling rate ($\Delta k$) in the diffraction patterns. When reconstructing using only the bright-field discs ($\beta \approx \alpha$), the resolution of ptychographic phases is observed to improve for samples that are thinner than ~6 nm as the sample thickness increases. It indicates that more high-frequency information is encoded in CBEDs by multiple scattering. Consistent with the results shown in FIG. 1, as the sample thickness increases, the difference in the resolution between the two collection angles decreases, indicating that electrons carrying higher frequency information enter the bright-field discs as multiple scattering enhances.

As the sample becomes thicker than ~6 nm, resolution in ptychographic phases decreases with sample thickness. In this regime, strengthened multiple scattering degrades the resolution. One of the possible reasons is that strong multiple scattering generates fine features in CBEDs, which vary more rapidly than sampling rate in reciprocal space (the detector plane). In this condition, the diffraction intensity will be under-sampled and the high-frequency information cannot be well decoded by ptychography. Therefore, we conclude that instead of using larger collection angle, reducing $\Delta k$ is more useful for achieving higher resolution for thick samples that generate strong multiple scattering.

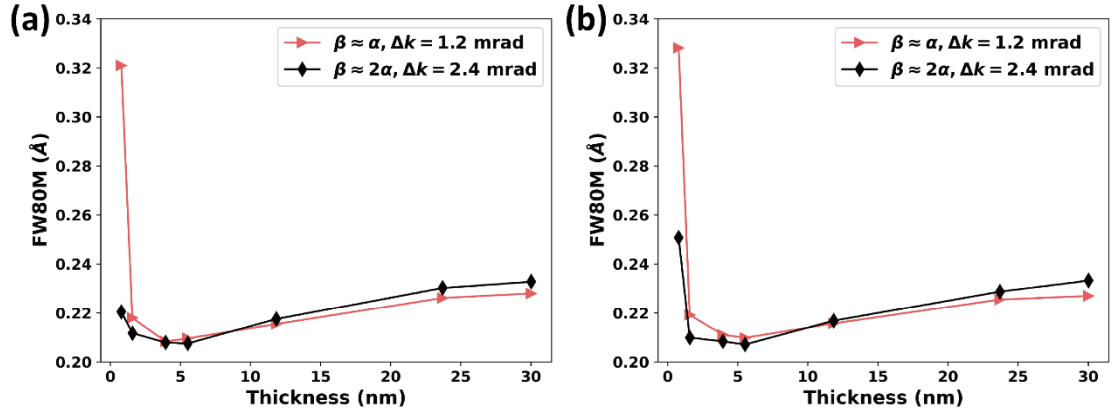

FIG. 3. FW80M of oxygen columns in phases recovered using datasets with different collection angles and reciprocal space sampling rates. Reconstructions with small and large collection angles are showed with red triangles and brown diamonds, respectively. (a) Noise-free datasets. (b) Poisson noise is added to the simulated datasets corresponding to the experimental dose value of $4\times10^6$ e/Å$^2$.

To summarize, the limiting factor for ptychographic resolution varies with sample thickness. For samples thinner than ~6 nm, ptychographic resolution is limited by the collection angle; larger collection angle results in higher resolution. For thicker samples, ptychographic resolution is limited by the sampling rate in diffraction patterns; fine sampling is required to take full use of the information encoded in the diffraction patterns. Recently, Gilgenbach et al. also suggest that relatively large Ronchigram magnification (i.e. fine sampling in diffraction patterns) leads to better convergence of multislice electron ptychography [31].

Electron dose is another important factor limiting the resolution of ptychography [11,32-35]. As shown in FIG. 4(b), the ptychographic resolution using large collection angle is more easily to be degraded by Poisson noise, especially for thin sample. This is as expected because high-angle scattering signal is much weaker than the signal in bright-field disk and thin samples generally scatter fewer electrons to high angle than thick samples. It is worth to have more detailed discussion about balancing sampling rate in diffraction space and collection angle under different magnitude of noise and multiple scattering.

To explain the enhancement effect of multiple scattering to the resolution of ptychography, we divide the electrons inside bright-field disc into two parts. Electrons in the first part have never been scattered outside the bright field during the multiple scattering process (yellow trajectory in right image in FIG. 5(a)), while those in the other part have been scattered to dark field but redirected to bright-field disc (green trajectory in right image in FIG. 5(a)). The two kinds of electrons are denoted as 'mBF' (maintained in bright field) and 'rBF' (redirected to bright field), respectively. We first separate the contributions of these two kinds of electrons to the bright-field disc intensity in simulation and then discuss their influence on the ptychography resolution.

In the multislice formalism of dynamic diffraction, the exit-wave $\varphi_{ext}^{BF}(r)$ in the bright-field disc is written as

$$\varphi_{ext}^{BF}(r) = \mathcal{P}\{\ldots\mathcal{P}\{\mathcal{P}\{\varphi_{in}O_1\}O_2\}O_3\ldots\}O_N, \tag{1}$$

$$\mathcal{P}\{A\} = \mathcal{F}^{-1}\{\mathcal{F}\{A\}(k) \cdot p(k)\} \tag{2}$$

Where $\varphi_{in}$ is the incident electron probe and $O_i$ is the $i$th slice of the object. $A$ stands for a 2D function. $\mathcal{P}$ stands for the near-field propagation described by the Fresnel propagator $p$. $\mathcal{F}$ is the Fourier transform. The exit-wave of 'mBF' electrons can be calculated with the following modification on the propagator:

$$\varphi_{ext}^{mBF}(r) = \mathcal{P}_m\{\ldots\mathcal{P}_m\{\mathcal{P}_m\{\varphi_{in}O_1\}O_2\}O_3\ldots\}O_N, \tag{3}$$

$$\mathcal{P}_m\{A\} = \mathcal{F}^{-1}\{\mathcal{F}\{A\}(k) \cdot m(k)p(k)\}, \tag{4}$$

$$m(k) = \begin{cases} 1, k \leq k_0\alpha \\ 0, k > k_0\alpha \end{cases} \tag{5}$$

where $k_0$ is the wave vector. Thus, the exit-wave of 'rBF' electrons is

$$\varphi_{ext}^{rBF}(r) = \varphi_{ext}^{BF}(r) - \varphi_{ext}^{mBF}(r), \tag{6}$$

Diffraction patterns are the intensity of the Fourier transform of exit-waves, which are shown in FIG. 5(b). With the sample thickness increases, $I^{rBF}$ strengthens while $I^{mBF}$ weakens. Also, the features in $I^{rBF}$ is much finer than those in $I^{mBF}$, indicating smaller detector pixels are needed to capture these features.

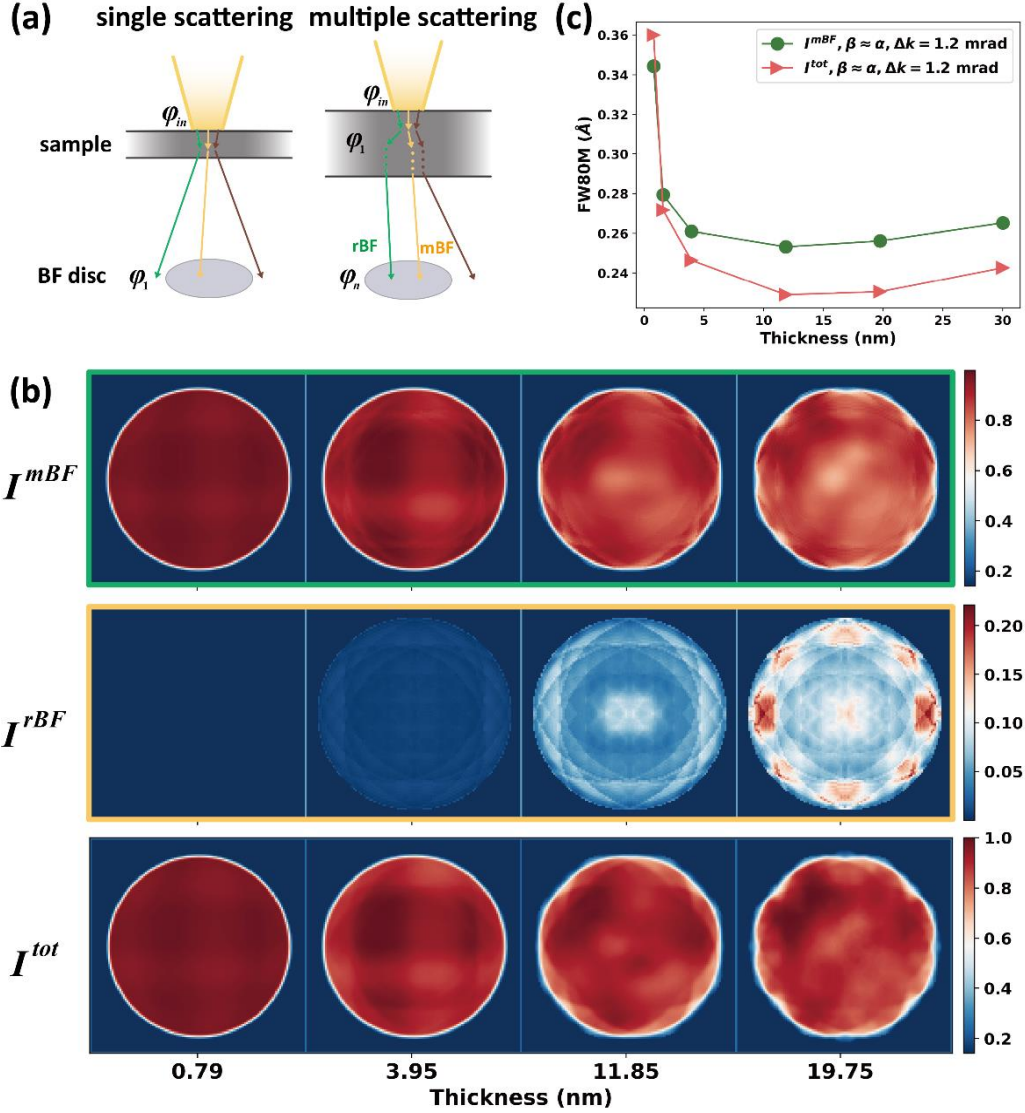

FIG. 5. Contributions of different parts of bright-field electrons to the diffraction intensity and reconstruction. (a) Schematics of kinematic (single) and dynamic (multiple) scattering. $\varphi_{in}$, $\varphi_1$ and $\varphi_n$ stand for incident electrons, electrons experienced once and n-times scattering, respectively. (b) PACBEDs of samples of different thickness. From top to the bottom, $I^{mBF}$, $I^{rBF}$ and $I^{BF}$. (c) FW80M of oxygen atomic columns in phases recovered using $I^{mBF}$ and $I^{BF}$.

Next, we investigate the influence of $I^{mBF}$ and $I^{rBF}$ on the ptychographic resolution. Reconstructions are done separately with $I^{mBF}$ and $I^{BF}$. When using $I^{mBF}$, propagation operator $\mathcal{P}_m$ is also adopted in the forward model of multislice reconstruction. FW80M of oxygen columns recovered based on $I^{mBF}$ are measured (shown with green circles in FIG. 5(c)) and compared with the reconstruction based on $I^{BF}$ (red triangles in FIG. 5(c)). As shown in FIG. 5(c), although $I^{rBF}$ is nearly one-order weaker than $I^{mBF}$, it obviously promotes the resolution. Considering fine features of $I^{rBF}$, fine sampling in reciprocal space is necessary to take full use of multiple scattering for higher resolution.

In conclusion, our research introduces a new approach to surpass the diffraction limit in electron microscopy. This method involves decoding high-frequency information, which is encoded in the bright-field discs through multiple scattering, using multislice ptychography. With the aid of multiple scattering, it is possible to achieve resolution beyond the diffraction limit, notably without relying on high-angle scattering.

This work was supported by the National Natural Science Foundation of China (52388201).